\documentclass[12pt,A4paper]{article}

\usepackage[T1]{fontenc}
\usepackage{amssymb}
\usepackage[dvips]{graphicx}
\usepackage[english]{babel}
\begin{document}

\section*{Proposal for new experimental schemes to realize the
Avogadro constant.}

\medskip

F.Biraben(1), M.Cadoret(1), P.Cladé(1), G.Genevès(2),
P.Gournay(2),S.Guellati-Khélifa(3), L.Julien(1), P.Juncar(3), E.
de Mirandes(1) and F.Nez(1).

\medskip

(1) Laboratoire Kastler Brossel, 4 place Jussieu, case 74, 75252,
Paris cedex 05, France.

\medskip

(2) Laboratoire National de Métrologie et d'Essais, ZA de
Trappes-Elancourt, 29 avenue Roger Hennequin, 78197 Trappes cedex,
France.

\medskip

(3) Institut National de Metrologie, Conservatoire National des
Arts et Métiers, 61, rue du Landy, 93210 La Plaine Saint Denis,
France.

\section*{Abstract}

We propose two experimental schemes to determine and so to realize
the Avogadro constant $N_{A}$ at the level of 10$^{-7}$ or better
with a watt balance experiment and a cold atom experiment
measuring $h/m(X)$ (where $h$ is the Planck constant and $m(X)$
the mass of the atom $X$). We give some prospects about achievable
uncertainties and we discuss the opportunity to test the existence
of possible unknown correction factors for the Josephson effect
and quantum Hall effect.

\section*{1. Introduction}

The best estimates of the fundamental constants are determined by
the Codata adjustments \cite{codata98,codata02}. A key weakness of
this adjustment is the lack of redundancy of input data.
Especially, this makes the determination of the Planck constant
$h$ and the Avogadro constant $N_{A}$ less confident
\cite{codata02}. In this paper, we propose to associate watt
balance and cold atoms experiments to define new direct
experimental ways to realize the Avogadro constant with a
competitive uncertainty in comparison with the best determination
obtained from the molar volume of a silicon crystal
(3.1$\times$10$^{-7}$) \cite{Na}. Moreover, the recent proposal of
redefinition of the kilogramme either by fixing the value of $h$
or the value of $N_{A}$ \cite{kg,cjb,kg2} will be reinforced by
many independent experimental determinations or comparisons of
these constants. Associated with a determination of $N_{A}$ at the
level of 10$^{-8}$, the proposed experiments can be used to check
the validity of the product $K_{J}^{2}R_{K}$ (where $K_{J}$ and
$R_{K}$ , the Josephson and the von Klitzing constants, are
respectively associated to the Josephson and the quantum Hall
effects).

\section*{2. Principle of the $N_{A}$ realization}

\subsection*{2.1 Quantities measured by the two experiments}

From its definition, $N_{A}$ can be expressed as the ratio between
atomic and macroscopic quantities such as the atomic mass and the
molar mass of any element. As it is already done in the above
mentioned single crystal silicon determination, any couple of
experiments giving access to these quantities may be considered.
Another possible combination consists in bringing together a watt
balance intended to link the kilogram to an invariant quantity and
a $h/m(X)$ experiment. Indeed, integrating these two experiments
leads to the determination of $N_{A}$ using the relation :

\begin{equation}
N_{A}=\frac{1}{h}\frac{h}{m(X)}A_{r}(X)M_{u}
\end{equation}

where $A_{r}$(X) is the relative atomic mass of $X$ and $M_{u}$ is
the molar mass constant ($M_{u}$=10$^{-3}$kg mol$^{-1}$).

The watt balance experiment consists in comparing a mechanical
power to an electromagnetic power \cite{kibble}. This comparison
is performed in two steps. In a static phase, the Laplace force on
a coil driven by a DC current and submitted to an induction field
is compared to the weight of a standard mass, linked to the
kilogram $M$. In a dynamic measurement, the voltage induced at the
terminals of the same coil is measured when it is moved in the
same field at a known velocity $V$. The measurement of electrical
quantities by comparison to the Josephson effect and the quantum
Hall effect (QHE) allows then to link the mass of the kilogram to
the product $K_{J}^{2}R_{K}$.

\begin{equation}
M K_{J}^{2}R_{K} = \frac{A}{g^{(w)}V}
\end{equation}

where $A=\frac{f_{1}f_{2}}{p}$  is proportional to the product of
the two Josephson frequencies involved in the voltage measurements
during the static and dynamic phases \cite{wb}. The dimensionless
$p$ is relative to the calibration of a resistance standard
against the quantum Hall effect and $g^{(w)}$ is the local
acceleration seen by the macroscopic mass $M$. Writing the
quantities to be measured in the experiment between brackets \{\},
the watt balance can determine \{$K_{J}^{2}R_{K}$\} if $g^{(w)}$
is measured independently with an absolute gravimeter, as well as
\{$K_{J}^{2}R_{K}g^{(w)}$\}.

The ratio $h/m(X)$ is determined by measuring the recoil velocity
($v_{r} = \hbar k / m(X)$) defined as the velocity induced by
light when an atom at rest absorbs a photon of momentum $\hbar k$.
The ratio $h/m(^{133}Cs)$ has been measured for the first time, at
Standford, using an atom interferometer with a relative
uncertainty of 15 ppb \cite{chu}. In another experiment, in Paris,
we have measured the ratio $h/m(^{87}Rb)$ with a relative
uncertainty of 13 ppb using Bloch oscillations in an optical
lattice \cite{clade}. A narrow velocity class is selected from
cold atoms sample with a Raman -pulse. This velocity class is
accelerated with Bloch oscillations. This process allows us to
transfer efficiently a high number of photon momenta
\cite{battesti}. The final velocity is measured with another Raman
 pulse. This experiment can run in two modes leading to two ways to
determine the Avogadro constant labelled $N_{A}^{(1)}$ and
$N_{A}^{(2)}$ hereafter.

In the first mode, a vertical optical standing wave is used to
hold the atoms against gravity \cite{stat}. The atoms oscillate at
the same place at the Bloch frequency ($\nu_{Bloch}$):

\begin{equation}
\nu_{Bloch}=\frac{m(^{87}Rb)g^{(a)}\lambda_{opt}}{2h}
\end{equation}

where $\lambda_{opt}$ is the wavelength of the optical wave and
$g^{(a)}$ is the local acceleration seen by the atoms. The
quantity \{$h/m(^{87}Rb) g^{(a)}$\} is measured in terms of
frequencies.

If the two experiments are brought close enough, the two local
accelerations $g^{(w)}$ and $g^{(a)}$ can be compared accurately
with relative gravimeters \cite{vitu}. Combining the quantities
measured by the two experiments leads to determine:

\begin{equation}
\{K_{J}^{2}R_{K}g^{(w)}\}\{\frac{h}{m(^{87}Rb)g^{(a)}}\}\{\frac{g^{(a)})}{g^{(w)}}\}
\end{equation}

In the second mode of the $h/m(^{87}Rb)$ experiment, the atoms are
accelerated up and down with Bloch oscillations. The resulting
differential measurement of $h/m^{87}Rb$ is independent of the
local acceleration $g$ \cite{clade}. Again the quantity
\{$h/m(^{87}Rb)$\} is measured in terms of frequencies. The
combination of the two experiments gives:

\begin{equation}
\{K_{J}^{2}R_{K}\}\{\frac{h}{m(^{87}Rb)}\}
\end{equation}

\subsection*{2.2 Determination of NA}

The values assigned by theory to $K_{J}$ and $R_{K}$ are :

\begin{equation}
K_{J}=\frac{2e}{h}
\end{equation}
and

\begin{equation}
R_{K}=\frac{h}{e^{2}}
\end{equation}

where $e$ is the elementary charge. The theoretical value of
$K_{J}^{2}R_{K}$ is then $4/h$.

If these relations are considered to be exact, the realization of
$N_{A}$ labelled $N_{A}^{(1)}$ and $N_{A}^{(2)}$ can be written as
:
\begin{equation}
N_{A}^{(1)}=\{\frac{K_{J}^{2}R_{K}g^{(w)}}{4}\}\{\frac{h}{m(^{87}Rb)g^{(a)}}\}\{\frac{g^{(a)})}{g^{(w)}}\}A_{r}(^{87}Rb)M_{u}
\end{equation}

\begin{equation}
N_{A}^{(2)}=\{\frac{K_{J}^{2}R_{K}}{4}\}\{\frac{h}{m(^{87}Rb)}\}A_{r}(^{87}Rb)M_{u}
\end{equation}

Notice that the realization of $N_{A}^{(1)}$ does not need the
knowledge of the absolute values of the local gravity g$^{(a)}$
and g$^{(w)}$ but only their relative values. In the case of
$N_{A}^{(2)}$, the knowledge of requires an absolute measurement
of g$^{(w)}$ (see eq. 2).

\subsection*{2.3 Test of $K_{J}^{2}R_{K}$}

However, proposing a new definition of the kilogram in terms of a
fundamental constant requires to lay on both theoretical and
experimental arguments. Even if the reproducibility of the quantum
Hall effect and the Josephson effect has been tested with a
relative uncertainty better than 1$\times$ 10$^{-10}$ under
various experimental conditions (material, temperature,...), there
is no experimental proof at this level that $K_{J}$ and $R_{K}$
are equal to their theoretical values \cite{jecke,tsai}. At
present, the only way to verify that $R_{K}$ is effectively equal
to $h/e^{2}$ consists in comparing its value obtained from an
experiment involving a QHE setup and the Lampard calculable
capacitor \cite{trapon} to those derived from other experiments,
such as determinations. The situation is similar for $K_{J}$
(assumed to be equal to $2e/h$) whose experimental knowledge is
issued, up to now, from the use of electrometers \cite{funck},
even if a determination can be deduced from the watt balance
experiment, provided resistance measurements are made in SI values
by comparison, for example to a Lampard calculable capacitor.

A test of this exactness has been done in the last Codata
adjustment, using the multivariate analysis. The inconsistencies
observed among certain input data have conducted to relax the
strict condition of equality between $R_{K}$ and $K_{J}$ and their
theoretical values. Two more adjusted constants $\epsilon_{J}$ and
$\epsilon_{K}$ describing unknown correction factors have been
added in the adjustment. The expressions of $K_{J}$ and $R_{K}$
then become:

\begin{equation}
K_{J}=\frac{2e}{h}(1+\epsilon_{J})
\end{equation}

and

\begin{equation}
R_{K}=\frac{h}{e^{2}}(1+\epsilon_{K})
\end{equation}

Therefore the relations issued from the two approaches proposed
for the experiment may be rewritten as:

 \begin{equation}
N_{A}^{(1)}=\{\frac{K_{J}^{2}R_{K}g^{(w)}}{4}\}\{\frac{h}{m(^{87}Rb)g^{(a)}}\}\{\frac{g^{(a)})}{g^{(w)}}\}\frac{A_{r}(^{87}Rb)M_{u}}
{(1+\epsilon_{J})^{2}(1+\epsilon_{K})}
\end{equation}

\begin{equation}
N_{A}^{(2)}=\{\frac{K_{J}^{2}R_{K}}{4}\}\{\frac{h}{m(^{87}Rb)}\}\frac{A_{r}(^{87}Rb)M_{u}}{(1+\epsilon_{J})^{2}(1+\epsilon_{K})}
\end{equation}

\section*{3. Discussion}

We introduce in this paragraph the present status of the different
uncertainties of these two possible realizations.

For the determination of $N_{A}^{(1)}$, the overall relative
standard uncertainty is presently limited by the uncertainty of
the quantity \{$\frac{h}{m(^{87}Rb)g^{(a)}}$\}. This ratio has
been measured in a preliminary experiment with a relative
uncertainty of about 10$^{-6}$ \cite{stat}, mainly due to
vibrations and collisions with the background vapor. These two
technical limitations can be overcome by using a more suitable
vacuum chamber where the Bloch oscillations take place and by
improving the vibration isolation. For example, 4000 Bloch
oscillations have been recently observed during 10s in the gravity
field \cite{ferra}. The other components of uncertainty are
smaller : the uncertainty of the quantity
\{$K_{J}^{2}R_{K}g^{(w)}$\} can be extrapolated from \cite{stein}
at the level of 4$\times$ 10$^{-8}$, the relative atomic mass of
rubidium $Ar(^{87}Rb)$ is known with an uncertainty better than
2$\times$ 10$^{-10}$ \cite{brad} and the gravity transfer can be
performed with an uncertainty of the order 1$\times$10$^{-9}$ if
the two experimental setups are close enough [REFERENCE ?].

The different contributions to the relative standard uncertainty
(u$_{r}$) of $N_{A}^{(2)}$ extrapolated from the different results
are listed in the following table. Different values of
$\epsilon_{J}$ and $\epsilon_{K}$, taking (or not) into account
some input data, are given in \cite{codata02}. We use here the
values $\epsilon_{J}$ and $\epsilon_{K}$ calculated with all the
input data :

\begin{center}
\begin{tabular}{|c|c|c|}\hline
Quantity & Value(uncertainty) & Ref.\\ \hline

[$h/m_{Rb}$]& 4.591 359 291 (61) 10$^{-9}$m$^{2}$s$^{-1}$ &
\cite{clade}\\ \hline

$4/[K_{J}^{2}R_{K}]$ & 6.626 069 01(34) 10$^{-34}$ Js &
\cite{stein}\\ \hline

$A_{r}(^{87}Rb)$  & 86.909 180 520 (15)u & \cite{brad}\\ \hline

$\epsilon_{J}$ & -126 (81) 10$^{-9}$ & \cite{codata02}\\ \hline
$\epsilon_{K}$ &  23 (19) 10$^{-9}$  & \cite{codata02}\\ \hline
\end{tabular}

\smallskip
Table 1
\smallskip

\end{center}

If it is assumed that there is no statistical significant evidence
that the basic relations for $K_{J}$ and $R_{K}$ are not exact
\cite{codata02}, $\epsilon_{J}$ and $\epsilon_{K}$ as well as
their uncertainties can be considered as equal to 0. Then, the
relative uncertainty on $N_{A}^{(2)}$ is at the level of
5.3$\times$ 10$^{-8}$.
\begin{center}
$N_{A}^{(2)}$ = 6.022 141 83 (33) 10$^{23}$ mol$^{-1}$
\end{center}
This values may be compared to the one issued from the silicium
\cite{Na} and to the one recommended by Codata \cite{codata02}:
\begin{center}
$N_{A}^{(Si)}$ = 6.022 135 3 (18)10$^{23}$ mol$^{-1}$

$N_{A}^{(Codata)}$ = 6.022 141 5 (10) 10$^{23}$ mol$^{-1}$
\end{center}
If we now consider the possible values of $\epsilon_{J}$ and
$\epsilon_{K}$, and the associated uncertainties issued from the
Cotata tests (see table 1), a new value of $N_{A}^{(2)}$ can be
determined :
\begin{center}
$N_{A}^{(2)}$= 6.022 143 21 (103) 10$^{23}$ mol$^{-1}$
\end{center}
The covariance factor between $\epsilon_{J}$ and $\epsilon_{K}$ is
extremely small as $\epsilon_{K}$ is determined mainly by the
measurements of and $R_{K}$, while $\epsilon_{J}$ should depend
only weakly on these measurements \cite{barry}. Taking into
account a null value for this covariance leads to a relative
uncertainty on $N_{A}^{(2)}$ of 2.8$\times$10$^{-7}$.

If, as mentioned by the groups in charge of the silicon project,
an uncertainty of 2$\times$10$^{-8}$ is expected in the future
\cite{beck}, the proposed determinations conduct to establish a
direct link between $R_{K}$ and $K_{J}$ derived from solid state
physics and $h$ derived from atomic physics. In that case,
gathering the two experiences could improve significantly our
confidence in the coherence of the phenomena on which a new
definition of the kilogram could be established and on the
experimental data on which its mise en pratique could be based.
Considering the above mentioned uncertainties, this could lead to
know the product $h K_{J}^{2}R_{K}/4$ with a relative uncertainty
of 5.3$\times$10$^{-8}$.

\section*{4. Conclusion}

We propose here new competitive schemes to realize the Avogadro
constant based on the conjunction of $h/m(^{87}Rb)$ experiment and
watt balance experiment where all quantities are measured in terms
of frequency. This proposal emphasizes the strong interest of
having a cold atom experiment nearby the watt balance. The
versatility of a cold atom experiment which can be used to measure
either $g^{(w)}$ \cite{cheinet} or $h/m(^{87}Rb)$ enables the
realization of $h$ and $N_{A}$ with the same watt balance.
Presently, provided it is considered that $K_{J}$ and $R_{K}$ are
equal to their theoretical value without any uncertainty, $N_{A}$
can be realized at a level of 5.3$\times$10$^{-8}$. This
uncertainty rises to 2.8$\times$ 10$^{-7}$ if one take into
account possible correction factors discussed by Codata 2002. This
shows that the knowledge on the von Klitzing and Josephson
constants may have a great influence on the numerical values of
fundamental constants and that any experiment to improve this
knowledge must be encouraged. The aim of the new experiment using
an enriched silicon sphere is to reach a 2$\times$10$^{-8}$
relative uncertainty for the determination of $N_{A}$. An
agreement at this level of uncertainty with the values of $N_{A}$
issued from the scheme proposed in this paper could then lead to a
test of the equality of the product $K_{J}^{2}R_{K}$ with $4/h$
with a relative uncertainty of 5.7$\times$10$^{-8}$. This strongly
emphasizes the interest of such determinations before the
redefinition of the kilogram and its \emph{mise en pratique}.

\end{document}